\documentclass{ifacconf}
\usepackage{subcaption}
\usepackage{svg}
\usepackage{graphicx} 
\usepackage{epsfig} 
\usepackage{amsmath} 
\usepackage{amsfonts}
\usepackage{amssymb}  
\allowdisplaybreaks
\usepackage{multirow}
\usepackage{multicol}
\usepackage{etoolbox} 

\usepackage{enumitem}

\usepackage{natbib}
\usepackage{float}
\usepackage{soul}
\usepackage{comment}
\usepackage{diagbox}
\usepackage{todonotes}

\usepackage{mathtools}
\newcommand{\R}{\mathbb{R}}
\newcommand{\C}{\mathcal{C}}

\newcommand{\K}{\mathcal{K}}
\newcommand{\mb}[1]{\mathbf{ #1 }}
\newcommand{\derp}[2]{\frac{\partial #1 }{\partial #2 }}

\theoremstyle{definition}
\newtheorem{definition}{Definition}

\newtoggle{ShowRemarks}
\togglefalse{ShowRemarks} 

 \iftoggle{ShowRemarks}{
\newcommand{\CRH}[1]{\textcolor{blue}{{\bf CRH:}  #1}}
\newcommand{\HX}[1]{\textcolor{red}{{\bf HX TODO:}  #1}}
\newcommand{\ISSUE}[1]{\textcolor{cyan}{{\bf ISSUE:}  #1}}
\newcommand{\TODO}[1]{\textcolor{Goldenrod}{{\bf TODO:}  #1}}
\newcommand{\UPDATE}[1]{\textcolor{red}{#1}}
}{
\newcommand{\CRH}[1]{}
\newcommand{\HX}[1]{}
\newcommand{\ISSUE}[1]{}
\newcommand{\TODO}[1]{}
\newcommand{\UPDATE}[1]{#1}
}

\begin{document}
\begin{frontmatter}
\title{Safe and Efficient \\ Data-driven Connected Cruise Control}

\author[footnoteinfo]{Haosong Xiao}
\author[footnoteinfo]{Chaozhe R. He}

\thanks[footnoteinfo]{H. Xiao and C. R. He are with the Department of Mechanical and Aerospace Engineering, University at Buffalo, Buffalo, NY, US. ${\tt\small \{haosongx, chaozheh\}@buffalo.edu}$}%

\begin{abstract}                
In this paper, we design a safe and efficient cruise control for the connected automated vehicle with access to motion information from multiple vehicles ahead via vehicle-to-vehicle (V2V) communication. 
Position and velocity data collected from a chain of human-driven vehicles are systematically leveraged to design a connected cruise controller that smoothly responds to traffic perturbations while maximizing energy efficiency. 
A safety filter derived from a control barrier function provides the safety guarantee. 
We investigate the proposed control design's energy performance against real traffic datasets and quantify the safety filter's energy impact.
It is shown that optimally utilizing V2V connectivity reduces energy consumption by more than 10\% compared to standard non-connected adaptive cruise control. 
Meanwhile, interesting interplays between safety filter and energy efficiency design are highlighted, revealing future research directions.
\end{abstract}

\begin{keyword}
Intelligent~Autonomous~Vehicles,~Control~Design,~Path~Planning~and~Motion~Control
\end{keyword}

\end{frontmatter}

\section{Introduction}
\vspace{-3mm}

Energy efficiency of vehicles is a critical metric in the automotive industry, since improving energy efficiency can bring great financial and societal benefits~\cite{ardalan2020book}. 
While improving vehicle powertrain designs provides fundamental energy efficiency improvement~\cite{ulsoy2012automotive},  vehicle operations also play an important role in energy consumption, and the large variations in driving behavior by human drivers could significantly compromise the energy efficiency~\cite{zarkadoula2007training}. 
Vehicle automation eliminates such variation, and extensive research has focused on optimizing the control input~(pedal, brake, and gear shift) to achieve the most efficient driving profile ~\cite{ardalan2020book, he2016pcc}.
Most of these studies assume no or ideal traffic conditions due to the lack of access to accurate and real-time traffic data, which is necessary for prediction and control. 

Vehicle-to-vehicle~(V2V) communication can potentially resolve this problem \cite{chaozhe2020fuel}: 
Peer-to-peer communication enables connected vehicles to share information for prediction and control, and facilitates cooperation among vehicles in traffic. 
Cooperative adaptive cruise control~(CACC) is a heavily researched design~\cite{barth2018reviewCACC} that leverages connectivity. 
Some seek to synchronize the speed of the platoon, guaranteeing string stability and maintaining desirable headway~\cite{VanDeWouW2014stringstb, yangzheng2018hinfPlatoon}, while others design predictive controllers that have access to the future motion plans of leading vehicles, thereby optimizing for energy optimal behaviors~\cite{borrelli2018CAVsurvey, Johansson_predictive_framework}. 
Despite the great benefit, the requirement for high penetration of both connectivity and automation will not be obtainable in the near future. 

Connected Cruise Control (CCC) \cite{orosz2016connected}, which focuses on using connectivity to benefit a single vehicle, 
requires a low penetration rate of automation, making it an appealing near-term solution. 
Theoretical and experimental results have shown great potential of CCC in improving energy efficiency \cite{Jin2018CCC, chaozhe2020fuel, minghao2022stochasticCCC}. 
One main gap in the application is that the existing energy-efficient CCC design does not provide any safety guarantees. 
The emerging control barrier function \cite{ames2019ECCTutorial, Alan2023CBFTutorial} provides a flexible and effective tool to certify and guarantee safety for existing controller design \cite{chaozhe2018}. \UPDATE{In this work, we seek to establish a real-time implementable framework for the efficient CCC design with rigorous safety guarantees.} 
More importantly, we aim to establish an implementable, energy-efficient CCC framework and quantify its robustness, as well as study the energy impact of the safety guarantee. 
Thus, our main contributions are
\begin{enumerate}
    \item Proposed an implementable, data-driven, safe, and efficient cruise control 
    for connected automated vehicles
    \item Established safety guarantees for energy-efficient connected cruise control design using control barrier function
    \item Present a reality check on the energy-saving performance while safety guarantee is enforced, analyze the robustness and impact of safety on energy efficiency
\end{enumerate}

\begin{figure*}[t]
    \centering
\includegraphics[width=0.98\linewidth]{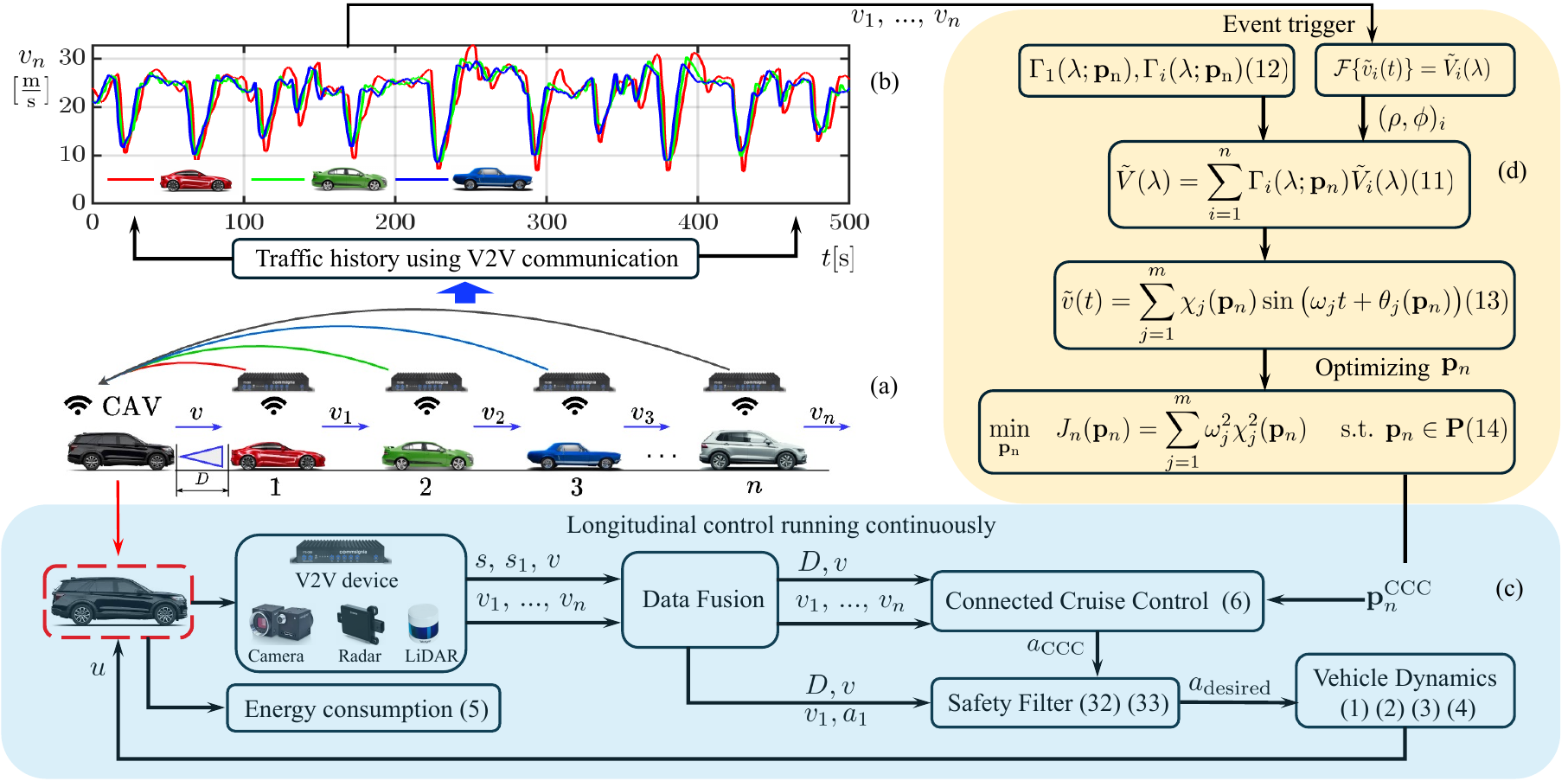}
    \caption{\small Design overview of the proposed data-driven safe and efficient connected cruise control framework. (a) Scenario of interest. (b) Traffic history samples acquired from V2V communication. (c) A longitudinal controller that operates continuously with a safety guarantee. (d) Event-triggered data-driven optimization on the connected cruise controller.}
    \label{fig:safeCCC}
\end{figure*}

\section{Design Overview}\label{sec:Safe_CCC_design}
\vspace{-2mm}
In this work, we focus on longitudinal controller design for a vehicle that travels on a straight, flat, single-lane highway section with dense and dynamic traffic; See Fig.~\ref{fig:safeCCC}(a). 
We assume the ego vehicle of interest (i.e., the black car at the end of the vehicle chain) is a connected automated vehicle (CAV) that can fuse the sensing results of vehicle ahead with standard sensors, with the information acquired from multiple vehicles ahead through connectivity on their position, speed, and acceleration. 
This allows us to collect data on traffic history. In Fig.~\ref{fig:safeCCC}(b), a 500-second sample is shown, where according to the velocity of three consecutive vehicles ahead of the CAV, the chain of vehicles has gone through multiple stop-and-go events, a typical scenario in daily driving. 

We present a data-driven, energy-efficient longitudinal design with a safety guarantee, which is summarized in Fig.~\ref{fig:safeCCC}(c,d). 
Detailed in Section \ref{sec:efficient_CCC}, the core connected cruise control (CCC) is a feedback controller that gives desired longitudinal acceleration and is parameterized with $\mb{p}_{n}$. 
To guarantee safety, a safety filter based on the control safety function is presented in Section~\ref{sec:safety_filter} that certifies safety and intervenes if necessary. 
Using a nominal model of vehicle dynamics, the lower-level command $u$ controls the CAV. 
The optimization of the CCC controller parameter $\mb{p}_{n}$ is formulated as one driven by collected traffic history data, which is also detailed in Section \ref{sec:efficient_CCC}.
Our design features two modules that operate on different schedules: the data-driven optimization of $\mb{p}_{n}$ is event-triggered and only executed when a data distribution shift occurs; the core CCC controller and the safety filter run continuously to ensure efficiency and safety.
We evaluate the energy and safety performance of the proposed design using real-world datasets in Section \ref{sec:results}.

\vspace{-2mm}
\section{Data-driven Efficient Connected Cruise Control Design}\label{sec:efficient_CCC}
\vspace{-2mm}
 In this section, we describe the data-driven, energy-efficient cruise control design that uses V2V information to control the longitudinal motion of the CAV. 

\begin{figure}
    \centering
    \includegraphics[width=0.52\textwidth]{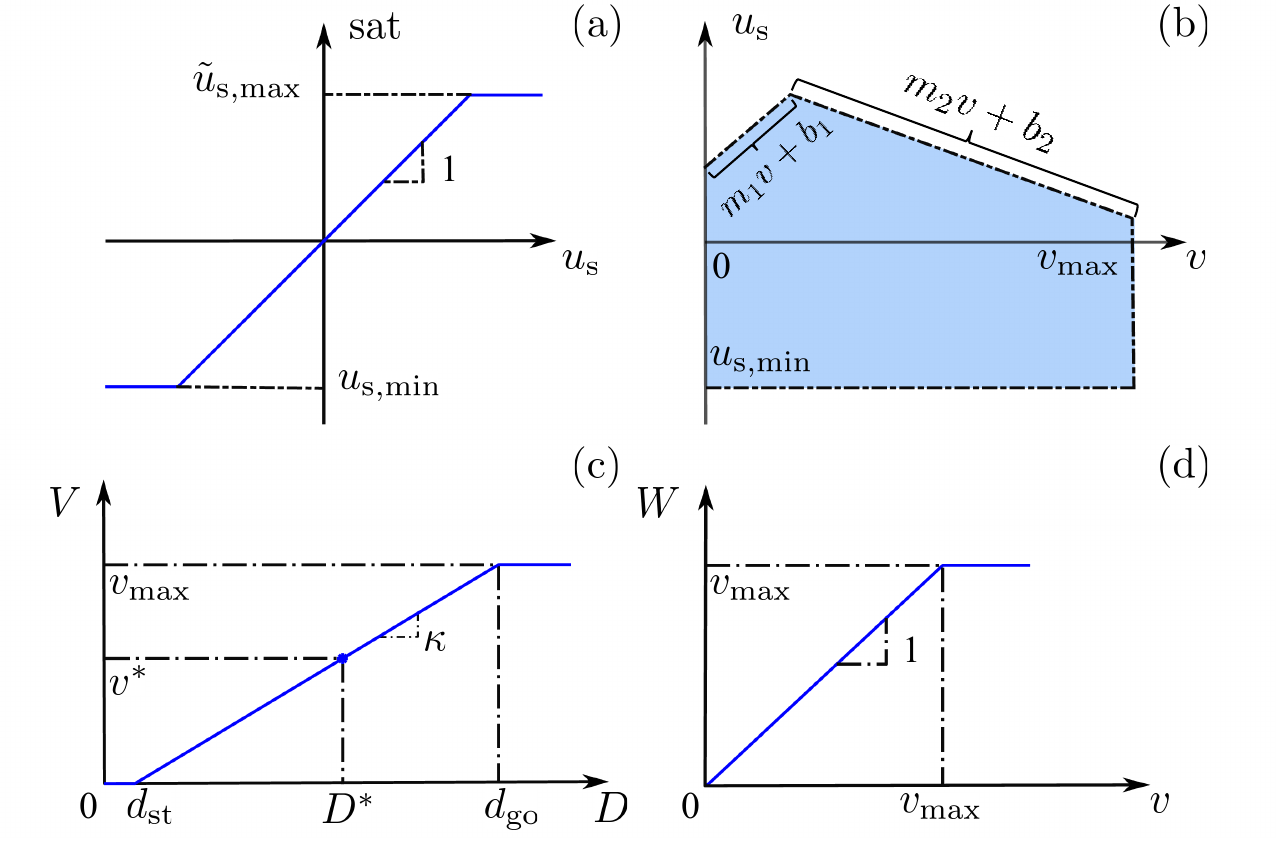}
    \caption{\small Nonlinear functions in the vehicle dynamics and CCC controller. 
             (a) Saturation function\,\eqref{eqn:sat_def}. (b) Acceleration limits\,\eqref{eqn:umax_def}.
             (c) Range policy\,\eqref{eqn:range_policy}. (d) Speed policy\,\eqref{eqn:speed_policy}.}
    \label{fig:sat_ovm_fun}
\end{figure}

We describe the longitudinal motion of the connected vehicle in the state space format as
\begin{equation}\label{eqn:cav_ss}
        \dot{D} = v_{1} - v~,\quad
        \dot{v} = -f(v) + {\rm sat}(u, v)~,\\
\end{equation}
where $D = s_{1} - s - l$ is the \emph{distance headway}, $s, v$ and $s_{1}, v_{1}$ denote the longitudinal position and speed of the CAV and its immediate predecessor, $l$ being the length of the CAV. $f(\cdot)$ is derived from longitudinal dynamics
\begin{equation}\label{eqn:f_def}
f(v) = \frac{1}{m_{\mathrm{eff}}}\left(m{\rm g}\zeta + kv^2\right).
\end{equation}
where $\rm g$ denotes the gravitational constant, $\zeta$~denotes the rolling resistance coefficient, and $k$ denotes the air resistance coefficient.
We model the powertrain to take a control command $u$ in the scale of acceleration originating from the powertrain torque (engine/motor) and the brake. 
Saturation function $\rm sat(\cdot, \cdot)$ represents powertrain limits in power/torque and braking capabilities:
\begin{subequations}\label{eqn:sat_0}
\begin{align}
\mathrm{sat}(u, v) &= \min\left\{\max\{u,u_{s,\min}\},\tilde{u}_{s, \max}(v)\right\}, \label{eqn:sat_def} \\
\tilde{u}_{s, \max}(v) &= \min\left\{u_{s, \max}, m_1 v + b_1, m_2 v + b_2\right\}, \label{eqn:umax_def}
\end{align}
\end{subequations}
as illustrated in Fig.\,\ref{fig:sat_ovm_fun}(a) and (b). 
In \eqref{eqn:sat_0}, $u_{s, \min}$ represents the minimum acceleration\,(i.e., maximum deceleration) due to the braking capability, and $u_{s, \max}$, $m_1$, $m_2$, $b_1$, $b_2$ are parameters determined by engine/motor power and torque limits.

To achieve a desired acceleration $a_{\rm d}$, a lower-level controller compensates the resistance force with a nominal model $\tilde{f}$ and delivers the final control $u$ to the vehicle \eqref{eqn:cav_ss}
\begin{equation}\label{eqn:ucomp_ftil}
u = \tilde{f}(v) + a_{\mathrm{d}}~.
\end{equation}
We remark that the compensation term may not be fully achieved because the saturation function \eqref{eqn:sat_def} cannot provide perfect compensation. This fact is accounted for in our simulation study.

Energy consumption is the main performance metric of the longitudinal control design in this work. It is evaluated with energy consumption per unit mass
\begin{equation}\label{eqn:w_def}
w = \int_{t_0}^{t_{\rm f}} v(t)g\Big(\dot{v}(t) + f(v(t))\Big)\mathrm{d}t~,
\end{equation}
where $g(x)=\max\{x, 0\}$ implies that braking does not consume or recover energy.
Note that the effects of energy recovery systems can be included by choosing different $g$ functions, but this is beyond the scope of this work.

We apply a connected cruise control (CCC) design to determine the desired acceleration of the CAV \cite{chaozhe2020fuel, minghao2022stochasticCCC}.
CCC design is inspired by the optimal velocity model~(OVM), which determines the desired acceleration $a_{\rm d}$ to follow the vehicle immediately ahead, but extends to incorporate up to $n-1$ vehicles that are beyond its line of sight:
\begin{equation}\label{eqn:ccc_nodelay}
a_{\rm CCC} =
\alpha\big(V(D)-v\big) + \sum_{i=1}^{n}\beta_i\big(W(v_i) - v\big)~.
\end{equation}
When $n=1$, the connected cruise control design is reduced to a standard non-connected adaptive cruise control (ACC) design. 
The range policy $V(D)$ determines the desired velocity for the distance headway $D$. 
The speed policy $W(v) = \min\{v_{\max}, v\}$ prevents the ego vehicle from speeding. 
We choose the range policy as
\begin{equation}\label{eqn:range_policy}
V(D) = \min\left\{v_{\max}, \max\{0, \kappa(D-d_{\mathrm{st}})\}\right\}~.
\end{equation}
As is shown in Fig.~\ref{fig:sat_ovm_fun}(c), when the distance headway is less than the stopping distance $d_{\rm st}$, the ego vehicle tends to stay still, while when the distance headway is larger than $d+ \tau_{\rm d} v_{\max}$, the ego vehicle intends to travel with maximum speed $v_{\max}$ without being influenced by the preceding vehicle. 
The desired velocity grows with constant gradient $1/\tau_{\rm d}$ where $\tau_{\rm d}$ is referred to as \emph{time headway}, which may be set differently to the trade-off between safety and traffic efficiency.
The speed policy 
\begin{equation}\label{eqn:speed_policy}
W(v_1) = \min\big\{v_{\max},v_1\big\}
\end{equation}
is used to prevent the CAV from speeding when the preceding vehicle goes faster than $v_{\max}$; see Fig.~\ref{fig:sat_ovm_fun}(d).

Given controller \eqref{eqn:ccc_nodelay}, our design goal is to find the optimal controller parameters (e.g., ${\mb p}_n=[\alpha, \kappa, \beta_1,\dots,\beta_n]$) such that the response generated by \eqref{eqn:cav_ss} minimizes the energy consumption \eqref{eqn:w_def}.
Building on our prior work \cite{chaozhe2020fuel}, we optimize the parameter $\mb p_{n}$ by leveraging traffic history data collected from V2V. 
Given a series of data $v_{i}(t)$, where $t\in[t_{0}, t_{\rm f}]$ with resolution of $\Delta t$, the velocity perturbation of vehicle $i$ can be described using the $m \leq \frac{t_{\rm f} - t_{0}}{2\Delta t}$ leading Fourier components \begin{equation}\label{eqn:input_inped}
\tilde{v}_{i}(t) = \sum_{j=1}^m \rho_{i,j}\sin(\omega_{j} t+\phi_{i,j})\,,
\end{equation}
where we discretized frequency $\omega_{j}=j\Delta \omega$, with $\Delta\omega=2\pi/(t_{\rm f}-t_{0})$.
Moreover, $\rho_{i,j}=\rho_{i}(\omega_{j})$ and $\phi_{i,j}=\phi_{i}(\omega_{j})$ are the amplitude and phase angle of speed oscillations at frequency $\omega_j$ for car $i$.

We use the linearized model of  (\ref{eqn:cav_ss},\ref{eqn:ccc_nodelay}) around the equilibrium given by the average speed of the traffic history data
 \begin{equation}\label{eqn:equilibrium}
 v(t)\equiv v_{i}(t)\equiv v^{\ast},\quad 
D(t) \equiv D^{\ast}, \quad v^{\ast}=V(D^{\ast})\,,
\end{equation}
for $i = 1,\dots,n$; cf.~(\ref{eqn:range_policy}) and Fig.~\ref{fig:sat_ovm_fun}(c).
We define $\tilde{v}_{i}$, $i=1,\ldots,n$ as the perturbations about the equilibrium velocities.
Assuming that the influence of the physical effects $f(v)$ can be negated by $\tilde{f}(v)$, the steady state response of the CAV in reaction to the historical traffic data can be expressed as 
\begin{equation}\label{eqn:TF_speed}
\tilde{V}(\lambda) = \sum_{i=1}^n \Gamma_i(\lambda;{\bf p}_n)\tilde{V}_{i}(\lambda)\,,
\end{equation}
where $\tilde{V}_i(\lambda)$ is the Laplace transform of the velocity perturbation $\tilde{v}_i(t)$.
The link transfer function from the $i$-th vehicle to the CAV can be written as
\begin{equation}\label{eqn:TF_j}
\begin{split}
  \Gamma_{1}(\lambda;{\bf p}_n)&=\frac{\alpha \kappa +\lambda \beta_{1}}{\lambda^{2}+\left(\alpha+\sum_{k=1}^{n}\beta_{k}\right)\lambda+\alpha \kappa }~,\\
  \Gamma_{i}(\lambda;{\bf p}_n)&=\frac{\lambda\beta_{i}}{\lambda^{2}+\left(\alpha+\sum_{k=1}^{n}\beta_{k}\right)\lambda+\alpha \kappa }~,
  \end{split}
\end{equation}
for $i=2,\ldots,n$. 
 Based on (\ref{eqn:input_inped},\ref{eqn:TF_speed},\ref{eqn:TF_j}), the steady-state oscillation of the CAV vehicle can be expressed as 
\begin{equation}\label{eqn:sinusoidal_speed_indepe}
  \tilde{v}(t)=\sum_{j=1}^{m} \chi_{j}({\bf p}_{n})\sin\big(\omega_{j} t +  \theta_{j}({\bf p}_n) \big)~,
\end{equation}
where the magnitude $\chi_{j}$ and phase angle $\theta_{j}$ can be derived using \eqref{eqn:TF_j} and therefore depend on ${\mb p}_n$.
To attenuate the transient response, we require the linearized dynamics
to be plant stable \cite{orosz2016connected}, which is achieved when all roots of the characteristic equation in \eqref{eqn:TF_j}
are located in the left half complex plane, leading to an admissible set for parameters ${\mb P} 
= \{\mb{p}_{n}~|~\alpha+\sum_{i=1}^{n}\beta_{i} > 0, \alpha >0, \kappa > 0\}$. 
Instead of using formulae 
(\ref{eqn:TF_j},\ref{eqn:sinusoidal_speed_indepe}) to construct the steady state response of the CAV to the signals given in \eqref{eqn:input_inped} and directly calculate the energy consumption using formula \eqref{eqn:w_def}, we draw insights from our prior work \cite{chaozhe2020fuel, minghao2022stochasticCCC} and acquire the CCC parameter by solving the following optimization problem
\begin{equation}\label{eqn:ObjFunTraf}
\text{Minimize} \quad
J_n({\bf p}_n) = \sum_{j=1}^m \omega_{j}^2 \chi_{j}^2({\bf p}_n)\,\quad {\rm s.t.} \,\mathbf{p}_{n} \in {\bf P}~.
\end{equation} 
This analytical, yet data-driven approach has a theoretical link to the original energy function \eqref{eqn:w_def}, and allows us to achieve robust energy-efficient performance against changes in traffic scenarios.
We note that the computational demand of such minimization is very low, as the cost function in \eqref{eqn:ObjFunTraf} does not require the reconstruction of the steady-state response in the time domain. 
\UPDATE{The summarized optimization process has been visualized in Fig.~\ref{fig:safeCCC}(b,d) and will be further explained mathematically from \eqref{eqn:ccc_nodelay} - \eqref{eqn:ObjFunTraf}.}

\vspace{-2mm}
\section{Safety Guarantee with Safety filter}\label{sec:safety_filter}
\vspace{-2mm}
In this section, we derive the safety filter used to provide a safety guarantee for the energy-efficient cruise control design \eqref{eqn:ccc_nodelay}. We will briefly review the control barrier function techniques and outline the safety filter design, referring to \cite{Alan2023CBFTutorial} for more details.

Consider the nonlinear control-affine system:
\begin{equation}
    \label{eqn:eom}
\dot{\mb{x}} = \mb{f}(\mb{x})+\mb{g}(\mb{x})\mb{u},
\end{equation}
with state ${\mb{x}\in\R^n}$, input ${\mb{u}\in\R^m}$, and continuous functions ${\mb{f}:\R^n\to\R^n}$ and ${\mb{g}:\R^n\to\R^{n\times m}}$. 
The input $\mb{u}$ is specified via a state-feedback controller ${\mb{k}:\R^n\to\R^m}$, 
yielding the closed-loop system dynamics:
\begin{equation}
    \label{eqn:cloop}
    \dot{\mb{x}} =   \mb{f}(\mb{x})+\mb{g}(\mb{x})\mb{k}_{\rm n}(\mb{x}).
\end{equation}
Such a controller, referred to as \emph{nominal controller}, meets some performance requirement, but not necessarily the safety requirements.
The system to which our data-driven CCC design introduced in Section \ref{sec:efficient_CCC} applies to, if assuming perfect compensation of $f(v)$,  can be expressed in the format of \eqref{eqn:eom} with
\begin{equation}\label{eqn:eom_cav}
\begin{split}
&\mb{x} = [v_{1}, D, v]^{\rm T},\quad \mb{u} = {\rm sat}(u, v)\\   
&\mb{f}(\mb{x}) = [a_{1}, v_{1} - v, 0]^{\rm T}, \mb{g}(\mb{x}) = [0, 0, 1]^{\rm T}
\end{split}    
\end{equation}
We note that \eqref{eqn:eom_cav} considers only the vehicle immediately ahead, which is the safety-critical one. 
The velocity of this vehicle is considered a state and its acceleration  $a_{1}$ is assumed to be an external signal, which could be time varying but is measurable and lower bounded: $a_{1} \geq \underline{a}_{1}$. The nominal controller $\mb{k}_{\rm n}(\mb{x})$ is given by (\ref{eqn:sat_0}, \ref{eqn:ucomp_ftil}, \ref{eqn:ccc_nodelay}).

Consider a set ${\C\subset \R^n}$ defined as the 0-superlevel set of a continuously differentiable function ${D:\R^n \to \R}$, yielding:
$ \C = \left\{\mb{x} \in \R^n : h(\mb{x}) \geq 0\right\} ,
    \partial\C = \{\mb{x} \in \R^n : h(\mb{x}) = 0\},
    {\rm Int}(\C) = \{\mb{x} \in \R^n : h(\mb{x}) > 0\} $
where $\partial\C$ and $\textrm{Int}(\C)$ are the \textit{boundary} and \textit{interior}, respectively, of the set $\C$. $\C$ is referred to as the \textit{safe set}, which we seek to stay forward invariant with respect to \eqref{eqn:eom}. While the nominal controller $\mb{k}_{\rm n}$ may not be able to achieve such a goal, control barrier functions (CBFs) may be used to enhance $\mb{k}_{\rm n}$ and provide a safety guarantee.
\begin{definition}[\textit{Control Barrier Function}, \cite{ames2017control}]
Let ${\C\subset\R^n}$ be the 0-superlevel set of a continuously differentiable function ${h:\R^n\to\R}$.
The function $h$ is a \textit{Control Barrier Function} (CBF) for the system \eqref{eqn:eom} on $\C$ if there exists ${\sigma\in\K_{\infty}^{\rm e}}$\footnote{A continuous function ${\alpha:\R\to\R}$ is said to belong to \textit{extended class $\cal{K}_\infty$} (${\alpha\in\cal{K}_{\infty}^{\rm e}}$) if ${\alpha(0)=0}$, $\alpha$ is strictly increasing, and ${\lim_{r\to\infty}\alpha(r)=\infty}$ and ${\lim_{r\to-\infty}\alpha(r)=-\infty}$.} such that for all ${\mb{x}\in\R^n}$:
\begin{equation}
\label{eqn:cbf}
    \sup_{\mb{u}\in\mathcal{U}}   \Bigg[  \overbrace{ \underbrace{ \derp{h}{\mb{x}}(\mb{x})\mb{f}(\mb{x})}_{L_\mb{f}h(\mb{x})}+\underbrace{\derp{h}{\mb{x}}(\mb{x})\mb{g}(\mb{x})}_{L_\mb{g}h(\mb{x})}\mb{u} 
    }^{\dot{h}(\mb{x},\mb{u})} \Bigg]  > -\sigma(h(\mb{x}))~.
\end{equation}
\end{definition}
\vspace{-2mm}
Given a CBF $h$ for \eqref{eqn:eom} on $\C$ and a corresponding function ${\sigma\in\cal{K}_{\infty}^{\rm e}}$, we can consider the point-wise set of all control values that satisfy \eqref{eqn:cbf}:
$K_{\rm CBF}(\mb{x}) = \{\mb{u}\in \mathcal{U} ~\left|~ \dot{h}(\mb{x},\mb{u})\geq-\sigma(h(\mb{x})) \right. \}.$ Given a valid CBF function, the goal of maintaining the performance of the nominal controller $\mb{k}_{\rm{n}}$ while ensuring the safety of the system \eqref{eqn:cloop} with respect to the set $\C$ can be achieved by an optimization-based safety-critical controller ${\mb{k}_{\rm QP}:\R^n\to\R^m}$ defined as:
\begin{align}
\label{eqn:SC-QP}
\small
\mb{k}_{\rm QP}(\mb{x}) =  \,\,\underset{\mb{u}\in\mathcal{U}}{\arg\min}  &  \quad \frac{1}{2} \| \mb{u}-\mb{k}_{\rm n}(\mb{x}) \|_2^2  \\
\mathrm{s.t.} \quad & \quad L_\mb{f}h(\mb{x}) +L_\mb{g}h(\mb{x})\mb{u}
\geq -\sigma(h(\mb{x})). \nonumber
\end{align}
This controller takes the same value as the nominal controller if the nominal controller meets the requirements for safety specified by the CBF $h$, i.e., ${\mb{k}_{\rm QP}(\mb{x})=\mb{k}_{\rm n}(\mb{x})}$ if ${\mb{k}_{\rm n}(\mb{x})\in K_{\rm CBF}(\mb{x})}$. If the nominal controller does not meet the safety requirements, i.e., $\mb{k}_{\rm n}(\mb{x})\notin K_{\rm CBF}(\mb{x})$, the input is chosen to meet the safety requirement with the smallest deviation from the value of $\mb{k}_{\rm n}$. Specifically, our prior work \cite{Alan2023CBFTutorial} has proven that for a single input ${(m=1)}$, if ${L_\mb{g}h(\mb{x}) < 0}$ for a particular $\mb{x}\in\R^n$, the controller \eqref{eqn:SC-QP} can be expressed in closed form as:
\begin{equation}
\label{eqn:switchstrucmin}
k_{\rm QP}(\mb{x}) = \min\left\{k_{\rm n}(\mb{x}), -\frac{L_\mb{f}h(\mb{x})+\sigma(h(\mb{x}))}{L_\mb{g}h(\mb{x})}\right\}.
\end{equation}
Similar cases may be derived for ${L_\mb{g}h(\mb{x}) > 0}$ cases and $L_\mb{g}h(\mb{x})$, with ${k_{\rm QP}(\mb{x})=k_{\rm n}(\mb{x})}$ when ${L_\mb{g}h(\mb{x})=0}$.

For our data-driven CCC design, we set the safety requirement as staying above a minimum time headway: $D -v \tau > 0$. Based on minimal stopping distance under safety critical cases, i.e., $a_{\rm d} = \underline{a}$ and $a_{1} = \underline{a}_{1}$, we synthesize the following CBF: 
\begin{equation}\label{eqn:ZCSF_compact}
h(D, v, v_1) = D-B(v,v_1)\geq 0,
\end{equation}
where $B$ is defined as follows.
If $\underline{a} \leq \underline{a}_1$ then
\begin{align}
\small
\overset{\text{if} ~~ \underline{a} \leq \underline{a}_1}{B(v,v_1)}=
 & \begin{cases}
   v\tau, & \text{if} \quad v_1\geq f_1(v),\\
    v\tau+\frac{(v-\underline{a}\tau)^2}{2\underline{a}}-\frac{v_1^2}{2\underline{a}_1},
    &  \text{if} \quad  v_1< f_1(v),\\
  \end{cases} \label{eqn:ZCSF_B_func} \\
\small 
\overset{\text{if} ~~ \underline{a} > \underline{a}_1}{B(v,v_1)}=
 & \begin{cases}
   v\tau,& \text{if} \quad v_1\geq f_{2}(v),\\
   v\tau + \frac{(v-\underline{a}\tau-v_1)^2}{2(\underline{a}-\underline{a}_1)},
   & \text{if} \quad f_{2}(v) < v_1< f_{3}(v),\\
   v\tau+\frac{(v-\underline{a}\tau)^2}{2\underline{a}}-\frac{v_1^2}{2\underline{a}_1},
   & \text{if} \quad v_1\geq f_{3}(v), \\
  \end{cases}\nonumber
\end{align}
where
\begin{equation}\label{eqn:bhat_a1>a_f23}
\small 
f_1(v)=\sqrt{\frac{\underline{a}_1}{{\underline{a}}}}(v-\underline{a}\tau), ~
  f_{2}(v)= v-\underline{a}\tau,~
  f_{3}(v)= \frac{\underline{a}_1}{\underline{a}}(v-\underline{a}\tau).\\
\end{equation}
The detailed derivation may be found in \cite{ames2017control,chaozhe2018}.
The existence of different cases is due to minimal distance headway occurring at different phases of the braking event: immediately after both vehicles apply the emergency brake, during the deceleration phase of the following vehicle, and at the end when the following vehicle comes to a complete stop.
By construction, \eqref{eqn:ZCSF_compact} is a valid control barrier function, and $h \geq B(v, v_1)$ implies that $D\geq v\tau$, which fulfills the safety requirement. 
We remark that \eqref{eqn:ZCSF_compact} admits a composition structure of 2 or 3 continuously differentiable functions, making it still satisfy the CBF definition \cite{Xu2017Correctness}. 

It can be verified that $L_{\mb g}h = - \partial B / \partial v < 0$ for all branches of $B$, thus, the safety filter that solves \eqref{eqn:SC-QP} can be summarized according to \eqref{eqn:switchstrucmin} as
\begin{equation}\label{eqn:Safety_filter}
    a_{\rm safe} = \min\{a_{\rm n}, a_{\rm cbf}\},
\end{equation}
where $a_{\rm n}$ is a nominal longitudinal acceleration before the safety filter (e.g. $a_{\rm CCC}$ (\ref{eqn:sat_0}, \ref{eqn:ucomp_ftil}, \ref{eqn:ccc_nodelay})) while
\begin{equation}\label{eqn:a_cbf}
\begin{split}
&a_{\rm cbf} = a_{\rm cbf}(D,v,v_1,a_1)
\\
&= \bigg(\frac{\partial B}{\partial v}\bigg)^{-1} \bigg(v_1-v-  \frac{\partial B}{\partial v_1}a_1+\gamma\big(D-B(v,v_1)\big)\bigg),
\end{split}
\end{equation}
with $\gamma > 0$ such that function as $\sigma(x) = \gamma x \in {\cal{K}_{\infty}^{\rm e}}$.

\begin{figure*}
    \centering
\includegraphics[width=0.96\linewidth]{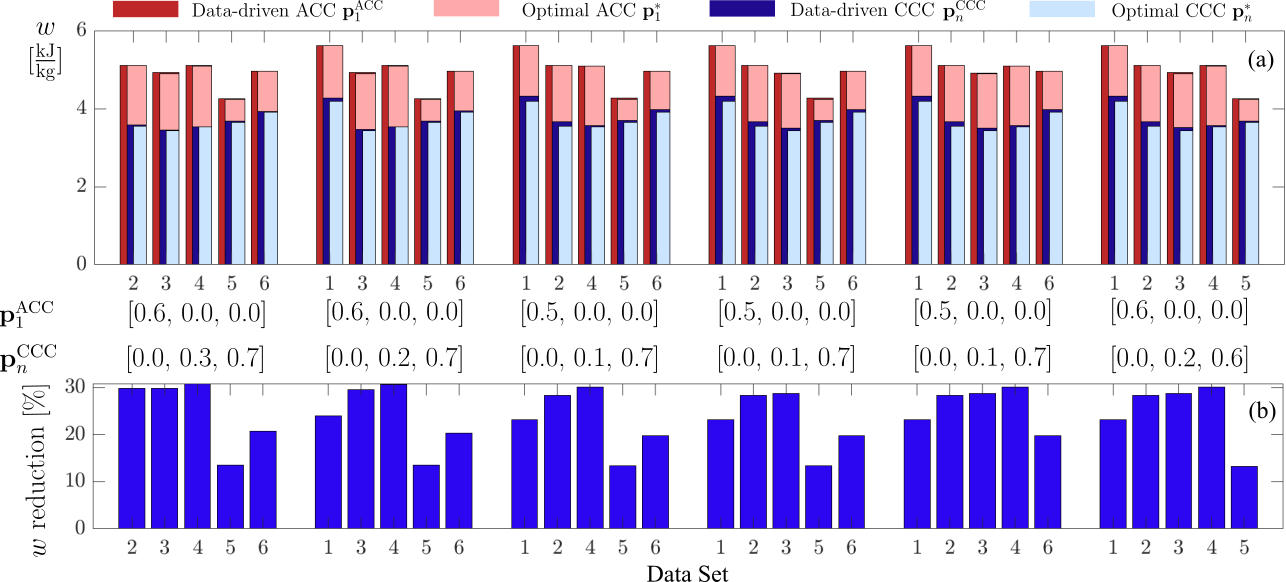}
\caption{\small Summary on the Energy Performance (a) Energy Consumption $w$ comparison among different test runs.(b) $w$ reduction percentage of data-driven CCC design $\mb{p}_{n}^{\rm CCC}$ over the data-driven ACC design $\mb{p}_{1}^{\rm ACC}$. Please note that $\mb{p}_{n}^{\rm CCC}$ and $\mb{p}_{1}^{\rm ACC}$ generated from different dataset can be the same (e.g. those for data set 3,4,5) due to the discrete search with step $0.1$ [1/s].} 

\label{fig:performance_summary}
\end{figure*}

\vspace{-2mm}
\section{Case Study}\label{sec:results}
\vspace{-2mm}
In this section, we consider the driving scenario shown in Fig.~\ref{fig:safeCCC}(a) and apply the proposed safe and efficient connected cruise control design summarized in Fig.~\ref{fig:safeCCC}(c,d) to a passenger vehicle. 
Through simulation, we evaluate the effectiveness of the proposed connected cruise control over non-connected cases.

We have recordings of six 500-second datasets with similar stop-and-go behavior by the human-driven vehicles (cf. Fig.~\ref{fig:safeCCC}(b)). 
For the details of how the data were collected, see \cite{Jin2018CCC}. We use the following event-trigger schedule to evaluate the proposed design: optimize $\mb{p}_{n}$ based on one dataset and test its performance over the other five datasets, resulting in six test runs of the proposed design. 
In each test run, we acquire the corresponding optimal CCC design $\mb{p}_{n}^{\rm CCC}$ based on one historical data set of $n$ vehicles ahead. To access the optimality gap, for each dataset, we also conduct a brute-force search among all the $\mb{p}_{n}$ (with a resolution of 0.1 [1/s] for each parameter in $[0, 2.0]$[1/s]) and find the $\mb{p}_{n}^{\ast}$ that minimize \eqref{eqn:w_def}, i.e., the global optimal parameter for each dataset. 
Correspondingly, we can acquire the baselines $\mb{p}_{\rm 1}^{\rm ACC}$ and $\mb{p}_{\rm 1}^{\ast}$ for non-connected cases using the information immediately ahead (c.f., $v_{1}$ in Fig.~\ref{fig:safeCCC}).
As a the case study we fixed $\alpha$ and $\kappa$ and optimize parameters of 3 vehicles ahead $\mb{p}_{3} = [\beta_{1}, \beta_{2}, \beta_{3}]$. The parameter used and code can be found online~\footnote{https://github.com/CHELabUB/safe\_resillent\_CCC}.  

\vspace{-3mm}
\subsection{Benefit of Connectivity}
\vspace{-3mm}
We summarize the performance of the proposed design over the six different test runs in Fig.~\ref{fig:performance_summary}. 
The 5 bars in each group in Fig.~\ref{fig:performance_summary}(a) correspond to the energy performance \eqref{eqn:w_def} when testing against datasets the design is not based on. Each bar shows four energy metric brought by the data-driven ACC design $\rm \mb{p}_{1}^{ACC}$ (red), the global optimal ACC design $ \mb{p}_{1}^{\ast}$ (light red), the data-driven CCC design $ \mb{p}_{n}^{\rm CCC}$ (blue) and the global optimal CCC design $\mb{p}_{n}^{\ast}$ (light blue).
In all six test runs, the proposed data-driven CCC $\mb{p}_{n}^{\rm CCC}$, which leverages beyond-line-of-sight information acquired from connectivity, can achieve better performance than the non-connected ACC design $\mb{p}_{1}^{\rm ACC}$. The percentage of reduction is summarized in Fig.~\ref{fig:performance_summary}(a). 
Furthermore, $\mb{p}_{n}^{\rm CCC}$ gives consistent improvement over global optimal ACC $\mb{p}_{1}^{\ast}$, and the gaps to the global optimal CCC $\mb{p}_{3}^{\ast}$ are small. Note that $\rm \mb{p}_{3}^{\ast}$ and $\mb{p}_{1}^{\ast}$ are not achievable in practice as they are acquired through brute force search. 

To reveal the reason for energy efficiency improvement, we plot the distance headway, speed, and $a_{\rm d}$ Fig.~\ref{fig:v_d_ad_profile_ACC_vs_CCC} for ACC (red dotted curves in the left column) and CCC (blue dash-dotted curves in the right column), respectively. 
In both cases, the CAV is following the vehicles ahead, traveling with profiles shown in Fig.~\ref{fig:safeCCC}(a).
The connectivity leads to milder variations in acceleration and deceleration, thereby resulting in milder speed variations. On the other hand, the distance headway is maintained in similar ranges between the ACC and CCC designs, showing comparable car-following performance.

\vspace{-3mm}
\subsection{Impact of Safety Filter on Energy Efficiency}
\vspace{-3mm}

As the safety filter \eqref{eqn:Safety_filter} always leads to a a desired acceleration smaller than the original CCC controller suggested, we are interested to see the energy-saving impact of the safety filter if any, i.e., whether it is improving the energy efficiency by mostly requesting milder acceleration, or hurting energy efficiency by demanding heavier brake. 
We further define the following metrics:
1) Energy wasted by using the brake:
\begin{equation}\label{eqn:w_brake}
    w_{\rm brake} = 
\int_{t_0}^{t_{\rm f}} v(t)g\Big(-\dot{v}(t) - f(v(t))\Big)\mathrm{d}t~;
    \end{equation}    
2) The time percentage of $h = D - B(v, v_1) <0$;
3) Safety violation margin
    \begin{equation}\label{eqn:h_margin}
        h_{\rm margin} = 
\int_{t_0}^{t_{\rm f}} g\Big(B(v, v_{1}) -D\Big)\mathrm{d}t~,
    \end{equation}
where $g(x)=\max\{x, 0\}$ for both cases.

\begin{figure*}
    \centering
\includegraphics[width=0.99\linewidth]{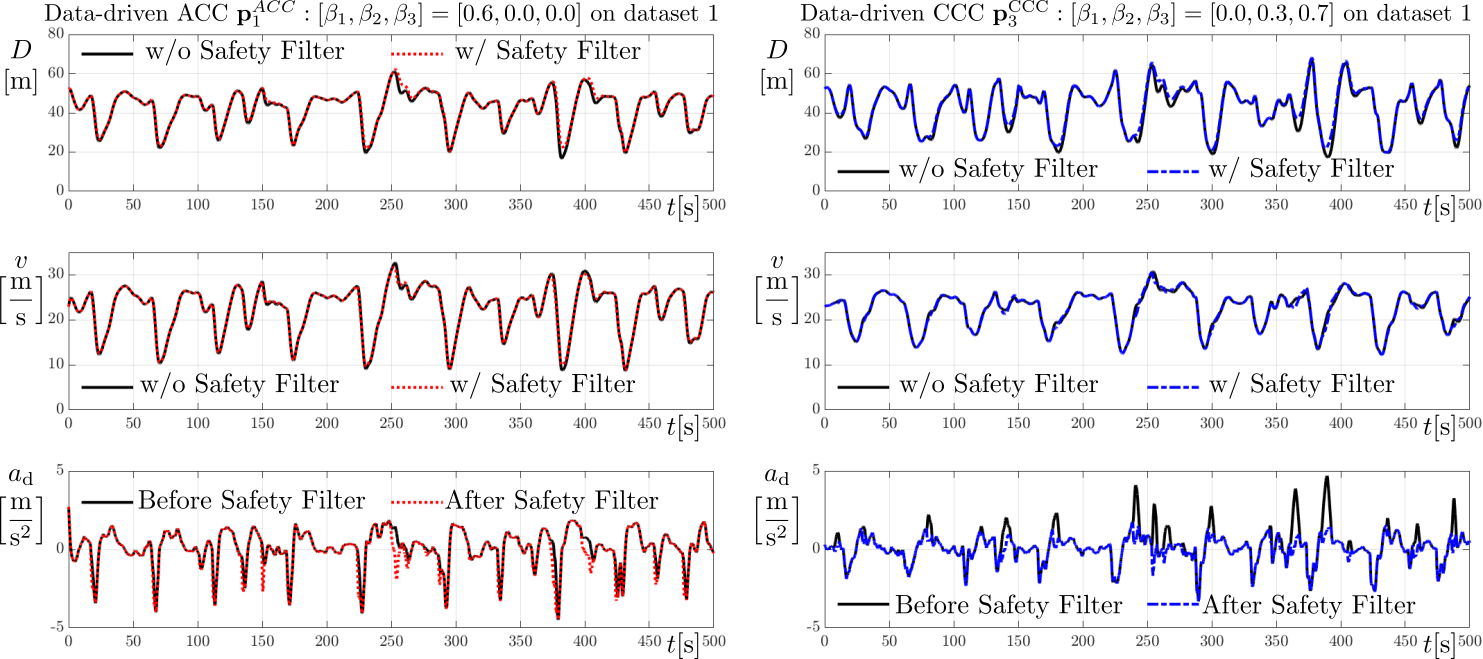}
    \caption{\small Profile comparison ((a) distance headway and (b) speed) between without Safety filter (black solid curves) and with safety filter for data-driven ACC (red dotted curve) and CCC (blue dash-dotted curves) designs. (c) Acceleration demand profile before (black) and after the safety filter for runs with the safety filter.} 
\label{fig:v_d_ad_profile_ACC_vs_CCC}
\end{figure*}

We found that while safety filters generally improve energy efficiency while providing a safety guarantee for ACC designs, the impact on CCC designs is mixed: safety filters do not always improve energy efficiency. We report the extreme case in our study where the energy efficiency of CCC was compromised the most with the safety filter, and summarize the metrics in Table~\ref{tab:CCC_vs_ACC_safety_filter}. 
The corresponding profiles for runs without safety filter \eqref{eqn:Safety_filter} are shown as black solid curves in Fig.~\ref{fig:v_d_ad_profile_ACC_vs_CCC}. 
While safety filter guarantees the safety (cf., $h<0$ percentage and $h_{\rm margin}$ drops to zero), it leads to less brake usage for the ACC design but more for the CCC design (c.f., an increase in $w_{\rm brake}$). 
According to the $a_{\rm d}$ profile, the acceleration request by CCC design encounters more intervention by the safety filter than that by the ACC design, indicating that better coordination between the CCC controller \eqref{eqn:ccc_nodelay} and safety filter \eqref{eqn:Safety_filter} may be attempted to increase energy efficiency further.

 \begin{table}
\centering
\begin{tabular}{|c|c|c|}
\hline
(w/o \eqref{eqn:Safety_filter}) / (w/ \eqref{eqn:Safety_filter}) & ACC & CCC \\
\hline
{$\mb{p}_{3} = [\beta_{1}, \beta_{2}, \beta_{3}]$} & {[0.6, 0.0, 0.0]} & {[0.0, 0.3, 0.7]} \\
\hline
{$w$ [${\text{kJ}}/{\text{kg}}$]} & {5.7845 / 5.6149} & {4.2040/4.2348} \\
\hline
$w_{\rm brake}$ [${\text{kJ}}/{\text{kg}}$] & {3.7931 / 3.6289} & {2.3120/2.3363} \\
\hline
{$h < 0$ percentage [\%]} & {10.52 / 0.00} & {15.82 / 0.00} \\
\hline
{$h_{\rm margin}$ [m]} & 0.5005/0.000 & 0.7839/0.0000 \\
\hline
\end{tabular}
\caption{\small Energy impact by Safety Filter}\label{tab:CCC_vs_ACC_safety_filter}
\vspace{-3mm}
\end{table}

\vspace{-2.5mm}
\section{Conclusions}\label{sec:concl}
\vspace{-2mm}
In this paper, we designed a safe and efficient cruise control for the CAV that has access to motion information from multiple vehicles ahead via V2V communication. 
Position and velocity data collected from a chain of human-driven vehicles are systematically leveraged to design a connected cruise controller that smoothly responds to traffic perturbations while maximizing energy efficiency. 
A safety guarantee is achieved using a safety filter derived using the control barrier function. 
We investigated the performance impact on real traffic data sets and quantified the safety filter's energy impact.
It was demonstrated that optimally utilizing V2V connectivity results in more than a 10\% reduction in energy consumption compared to standard non-connected adaptive cruise control. 
Meanwhile, interesting interplays between safety filter and energy efficiency design are highlighted, revealing future research directions.
For future work, we are investigating how to incorporate a speed profile preview into the safety filter design to allow more proactive intervention tailored to improve energy efficiency. We are also working on schedule design for CCC parameter optimization and extending the system with delay and a more dynamic communication structure.

\vspace{-3mm}


\end{document}